\documentclass[letterpaper]{article}
\usepackage{spconf,amsmath,graphicx}
\usepackage{lineno}
\usepackage{bm}
\usepackage{mathrsfs}
\usepackage{amsmath,cases}
\usepackage{amssymb}
\usepackage{textcomp}
\usepackage{graphicx}
\usepackage{epstopdf}
\usepackage{xcolor}
\usepackage{geometry}
\newgeometry{left=0.75in,top=0.96in,right=0.75in,bottom=0.98in}
\title{Study of filtered-x logarithmic recursive least $p$-power algorithm}

\name{Zongsheng Zheng$^{*}$, Lu Lu$^{**}$, Yi Yu$^{\star}$, Rodrigo C.~de Lamare$^{\dagger\ddagger}$, Zhigang Liu$^{\star\star}$
    \thanks{This work was partially supported by the National Natural Science Foundation of China (Nos. 61901285 and 61901400), Sichuan Science and Technology Fund under Grant 20YYJC3709, the Doctoral Research Fund of Southwest University of Science and Technology in China (No. 19zx7122), China Postdoctoral Science Foundation under Grant 2020T130453, the Fundamental Research Funds for the Central Universities under Grant 2021SCU12063, and Sichuan University Postdoctoral Interdisciplinary Innovation Fund.}
}

\address{
    * \normalsize{\emph{College of Electrical Engineering, Sichuan University, Chengdu 610065, China}}\\
    ** \normalsize{\emph{School of Electronics and Information Engineering, Sichuan University, Chengdu 610065, China}}\\
    $\star$ \normalsize{\emph{School of Information Engineering, Robot Technology Used for Special Environment Key Laboratory of Sichuan Province}}, \\
    \normalsize{\emph{Southwest University of Science and Technology, Mianyang 621010, China}}\\
    $\dagger$ \normalsize{\emph{CETUC, PUC-Rio, Rio de Janeiro 22451-900, Brazil}}\\
    $\ddagger$ \normalsize{\emph{Department of Electronic Engineering, University of York, York YO10 5DD, U.K.}}\\
    $\star\star$ \normalsize{\emph{School of Electrical Engineering, Southwest Jiaotong University,  Chengdu 610031, China}}\\
    \normalsize{zongsheng56@126.com, lulu19900303@126.com, yuyi\_xyuan@163.com, rcdl500@ohm.york.ac.uk, liuzg\_cd@126.com
}}
\begin{document}
\ninept
\maketitle

\begin{abstract}
For active impulsive noise control, a filtered-x recursive least $p$-power (FxRLP) algorithm is proposed by minimizing the weighted summation of the $p$-power of the \emph{a posteriori} errors. Since the characteristic of the target noise is investigated, the FxRLP algorithm achieves good performance and robustness. To obtain a better performance, we develop a filtered-x logarithmic recursive least $p$-power (FxlogRLP) algorithm which integrates the $p$-order moment with the logarithmic-order moment. Simulation results demonstrate that the FxlogRLP algorithm is superior to the existing algorithms in terms of convergence rate and noise reduction.
\end{abstract}
\section{Introduction}

On the basis of wave-superposition principle, active noise control
(ANC) is realized by utilizing adaptive filters to generate a signal
with same magnitude but opposite phase of the signal to be
canceled\cite{elliott1993active,tan2001adaptive,sicuranza2011generalized,lu2019power,lu2021survey,lu2020survey}.
Thanks to the low computational complexity and simple structure, the
filtered-x least mean square (FxLMS) algorithm is frequently
performed in the adaptive filtering algorithms
\cite{jidf,spa,intadap,mbdf,jio,jiols,jiomimo,sjidf,ccmmwf,tds,mfdf,l1stap,mberdf,jio_lcmv,locsme,smtvb,ccmrls,dce,itic,jiostap,aifir,ccmmimo,vsscmv,bfidd,mbsic,wlmwf,bbprec,okspme,rdrcb,smce,armo,wljio,saap,vfap,saalt,mcg,sintprec,stmfdf,1bitidd,jpais,did,rrmber,memd,jiodf,baplnc,als,vssccm,doaalrd,jidfecho,dcg,rccm,ccmavf,mberrr,damdc,smjio,saabf,arh,lsomp,jrpaalt,smccm,vssccm2,vffccm,sor,aaidd,lrcc,kaesprit,lcdcd,smbeam,ccmjio,wlccm,dlmme,listmtc,smcg},
As compared to the FxLMS algorithm, the filtered-x recursive least
square (FxRLS) algorithm \cite{kuo1996active} can provide faster
convergence rate at the cost of higher computational complexity.

Note that the FxRLS algorithm is based on the assumption that the
reference signal follows the Gaussian distribution. But in fact, the
target noise always contains outliers, that is, the impulsive noise
exists in the target
noise\cite{sun2006adaptive,akhtar2009improving,li2013active,yu2020diffusion}.
For such situations, the FxRLS algorithm always shows bad
performance and has stability problems. To address this problem,
instead of the mean-square error (MSE) used in the FxLMS algorithm,
the filtered-x least mean $p$-power (FxLMP) algorithm was proposed
\cite{leahy1995adaptive} by minimizing the mean $p$-power of error.
In \cite{wu2010active}, the filtered-x logarithmic least mean square
(FxlogLMS) algorithm was proposed by minimizing the squared
logarithmic transformation of the error signal. Simulation results
demonstrated that the FxlogLMS algorithm has good robustness and
better performance. Furthermore, to achieve faster convergence rate,
the filtered-x logarithmic recursive least squares (FxlogRLS)
algorithm was proposed \cite{wu2015recursive} by minimizing the
weighted summation of the logarithmic transformation of the \emph{a
posteriori} errors.

In this work, we try to work out the above problem in another way.
To investigate the characteristic of the target noise, a filtered-x
recursive least $p$-power (FxRLP) algorithm is proposed by
minimizing the weighted summation of the $p$-power of the \emph{a
posteriori} errors. Then, by integrating the $p$-order moment with
the logarithmic-order moment, we develop a filtered-x logarithmic
recursive least $p$-power (FxlogRLP) algorithm.

The main contributions of this paper are summarized as follows:

\noindent1) By defining a new cost function based on the $p$-power
of error, a new adaptive filtering algorithm is proposed, which is
an extension of the FxLMP algorithm in the recursive least squares
structure.

\noindent2) By integrating the $p$-order moment with the logarithmic-order moment, a novel adaptive filtering algorithm is developed. It is worth noting that this paper for the first time combines the benefits of the FxLMP-type algorithm and the FxlogLMS-type algorithm.

\noindent3) The properties of the proposed algorithms and their relationship to the existing algorithms are illustrated in detail.

The rest of this paper is organized as follows. Section 2 presents the derivation of the proposed FxRLP algorithm, and the proposed FxlogRLP algorithm is derived in Section 3. In Section 4, the mean stabilities of the proposed algorithms are analyzed. Section 5 illustrates the simulation results, and the conclusions are drawn in Section 6.

\section{FxRLP algorithm}
\begin{figure} [htbp]
    \centering
    \includegraphics[scale=0.62] {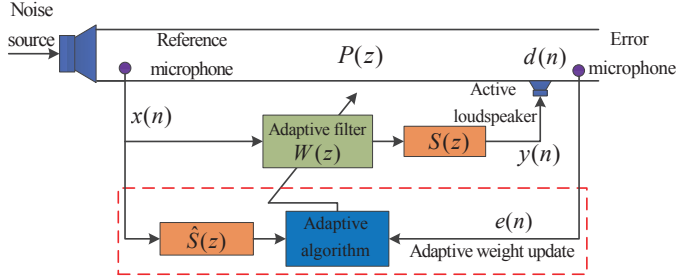}
    \caption{Block diagram of a single-channel feed-forward ANC system}
    \label{ANC}
\end{figure}

Fig. \ref{ANC} shows a single-channel feed-forward ANC system, where $P(z)$ represents the primary path between the reference signal $x(n)$ and the error microphone $e(n)$, $d(n)$ is the primary noise, $S(z)$ denotes the secondary path from the adaptive filter $W(z)$ to the error microphone $e(n)$, $\hat{S}(z)$ can be obtained by using an off-line or on-line system identification approach \cite{samarasinghe2016recent,yin2018functional,aslam2019robust}, and $y(n)$ represents the output of the secondary sound source. Then, the error microphone $e(n)$ can be calculated by
\begin{equation}
    e(n)=d(n)-y(n)=d(n)-s(n)*[{{\bm{w}}^{T}}(n)\bm{x}(n)]
\end{equation}
where $\bm{x}(n)={{[x(n),x(n-1),...,x(n-L+1)]}^{T}}$ represents the reference signal vector, $s(n)$ denotes the impulse response of $S(z)$, * represents the discrete convolution operator, and superscript $T$ denotes transposition.

\noindent\emph{Remark 1:} Since the impulsive noise can be modeled as a $\alpha $-stable process ($0<\alpha<2$, $\alpha$ is the characteristic exponent), the second-order moment of the impulsive noise is not finite and the adaptive algorithm based on the MSE suffers from the stability problem. Hence, it is wise to use the cost function with the fractional lower order moment ($p$-power of error, $1<p<\alpha $) for designing the robust adaptive algorithm.
$\hfill\blacksquare$

Here, the cost function of the FxRLP algorithm is defined as
\begin{equation}
    {{J}_\text{FxRLP}}(n)=\sum\limits_{i=1}^{n}{{{\lambda }^{n-i}}{{\left| \varepsilon (i,n) \right|}^{p}}}
\end{equation}
where $0\ll \lambda <1$ denotes the forgetting factor, $\varepsilon (i,n)=d(i)-{{\bm{w}}^{T}}(n){{\bm{x}}_{s}}(i)$ represents the \emph{a posteriori} error, $\bm{w}(n)$ is the adaptive filter tap weight vector with length $L$, ${{\bm{x}}_{s}}(n)=[{{x}_{s}}(n),{{x}_{s}}(n-1),...,{{x}_{s}}(n-L+1)]^{T}$, ${{x}_{s}}(n)=\hat{s}(n)*x(n)$, and  $\hat{s}(n)$ denotes the impulse response of $\hat{S}(z)$.

Taking the gradient of ${{J}_\text{FxRLP}}(n)$ with respect to $\bm{w}(n)$ and equating the result to zero, one gets
\begin{equation}
    \frac{\partial {{J}_\text{FxRLP}}(n)}{\partial \bm{w}(n)}=-p\sum\limits_{i=1}^{n}{{{\lambda }^{n-i}}\frac{{{\left| \varepsilon (i,n) \right|}^{p}}}{{{\left| \varepsilon (i,n) \right|}^{2}}}\varepsilon (i,n){{\bm{x}}_{s}}(i)}=0
\end{equation}
which can be written as
\begin{equation}
    \sum\limits_{i=1}^{n}{{{\lambda }^{n-i}}v(i,n){{\bm{x}}_{s}}(i)\bm{x}_{s}^{T}(i)\bm{w}(n)}=\sum\limits_{i=1}^{n}{{{\lambda }^{n-i}}v(i,n){{\bm{x}}_{s}}(i)d(i)}
    \label{grad}
\end{equation}
where $v(i,n)=\frac{{{\left| \varepsilon (i,n) \right|}^{p}}}{{{\left| \varepsilon (i,n) \right|}^{2}}}$.

From (\ref{grad}), we obtain the following expression for $\bm{w}(n)$
\begin{equation}
    \bm{w}(n)=\bm{P}(n)\bm{\theta}(n)
    \label{ww1}
\end{equation}
where
\begin{equation}
    \bm{P}(n)={{\bm{R}}^{-1}}(n),
\end{equation}
\begin{equation}
    \bm{R}(n)=\sum\limits_{i=1}^{n}{{{\lambda }^{n-i}}v(i,n){{\bm{x}}_{s}}(i)\bm{x}_{s}^{T}(i)},
\end{equation}
and
\begin{equation}
    \bm{\theta}(n)=\sum\limits_{i=1}^{n}{{{\lambda }^{n-i}}v(i,n){{\bm{x}}_{s}}(i)d(i)}.
\end{equation}

To derive an on-line algorithm, the following approximations are used:

\begin{equation}
\begin{aligned}
\bm{R}(n)&\approx\sum\limits_{i=1}^{n}{{{\lambda }^{n-i}}v(i,i){{\bm{x}}_{s}}(i)\bm{x}_{s}^{T}(i)}\\
&=\lambda \bm{R}(n-1)+v(n,n){{\bm{x}}_{s}}(n)\bm{x}_{s}^{T}(n),
\end{aligned}
\end{equation}
and
\begin{equation}
\bm{\theta}(n)\approx\sum\limits_{i=1}^{n}{{{\lambda }^{n-i}}v(i,i){{\bm{x}}_{s}}(i)d(i)}=\lambda \bm{\theta}(n-1)+v(n,n){{\bm{x}}_{s}}(n)d(n).
\label{zz1}
\end{equation}

By using Woodbury's matrix inversion lemma \cite{zhang2017variable,yu2019dcd,lu2020recursive}, the inverse autocorrelation matrix can be updated by
\begin{equation}
    \bm{P}(n)={{\lambda }^{-1}}\bm{P}(n-1)-{{\lambda }^{-1}}\bm{K}(n)\bm{x}_{s}^{T}(n)\bm{P}(n-1)
    \label{pp1}
\end{equation}
where the gain vector is defined as
\begin{equation}
    \bm{K}(n)=\frac{v(n,n)\bm{P}(n-1){{\bm{x}}_{s}}(n)}{\lambda +v(n,n)\bm{x}_{s}^{T}(n)\bm{P}(n-1){{\bm{x}}_{s}}(n)}.
    \label{k11}
\end{equation}

According to (\ref{pp1}), (\ref{k11}) can be simplified as
\begin{equation}
    \bm{K}(n)=v(n,n)\bm{P}(n){{\bm{x}}_{s}}(n).
    \label{k21}
\end{equation}

From (\ref{ww1}), (\ref{zz1}), (\ref{pp1}) and (\ref{k21}), $\bm{w}(n)$ can be updated by
\begin{equation}
    \begin{aligned}
         \bm{w}(n)=&\bm{P}(n)(\lambda \bm{\theta}(n-1)+v(n,n){{\bm{x}}_{s}}(n)d(n)) \\
         =&\lambda \bm{P}(n)\bm{\theta}(n-1)+v(n,n)\bm{P}(n){{\bm{x}}_{s}}(n)d(n) \\
         =&\bm{P}(n-1)\bm{\theta}(n-1)-\bm{K}(n)\bm{x}_{s}^{T}(n)\bm{P}(n-1)\bm{\theta}(n-1)\\&+\bm{K}(n)d(n) \\
         =&\bm{w}(n-1)+\bm{K}(n)\varphi (n) \\
    \end{aligned}
    \label{update1}
\end{equation}
where $\varphi (n)=d(n)-\bm{x}_{s}^{T}(n)\bm{w}(n-1)$ is the \emph{a priori} error.

\section{FxlogRLP algorithm}
According to the zero-order statistics \cite{gonzalez2006zero} which states that the $\alpha $-stable process is a logarithmic-order process with finite logarithmic moments, we define a new cost function as follows
\begin{equation}
    {{J}_\text{FxlogRLP}}(n)=\sum\limits_{i=1}^{n}{{{\lambda }^{n-i}}{{\log }^{p}}(1+\left| \varepsilon (i,n) \right|)}.
\end{equation}

Taking the gradient of ${{J}_\text{FxlogRLP}}(n)$ with respect to $\bm{w}(n)$ and letting the result to zero, we obtain
\begin{equation}
\begin{aligned}
    \frac{\partial {{J}_\text{FxlogRLP}}(n)}{\partial \bm{w}(n)}&=-p\sum\limits_{i=1}^{n}{{{\lambda }^{n-i}}\frac{{{\log }^{p-1}}(1+\left| \varepsilon(i,n) \right|)}{(1+\left| \varepsilon(i,n) \right|)\left| \varepsilon(i,n) \right|}\varepsilon (i,n){{\bm{x}}_{s}}(i)}\\
    &=0
\end{aligned}
\end{equation}
which can be rewritten as
\begin{equation}
    \sum\limits_{i=1}^{n}{{{\lambda }^{n-i}}v(i,n)\bm{x}(i){{\bm{x}}^{T}}(i)\bm{w}(n)}=\sum\limits_{i=1}^{n}{{{\lambda }^{n-i}}v(i,n)\bm{x}(i)d(i)}
    \label{grad2}
\end{equation}
where $v(i,n)=\frac{{{\log }^{p-1}}(1+\left| \varepsilon(i,n) \right|)}{(1+\left| \varepsilon(i,n) \right|)\left| \varepsilon(i,n) \right|}$.

The following derivation of the FxlogRLP algorithm is the same as that of the FxRLP algorithm, as shown in (\ref{ww1})-(\ref{update1}).

\noindent\emph{Remark 2:} To implement the FxRLP and FxlogRLP algorithms recursively and efficiently, we replace the \emph{a priori} error $\varphi (n)$ and the \emph{a posteriori} error $\varepsilon (n)$ by the residual noise $e(n)$ which can be measured by the error sensor. Here, $\varepsilon (n)$ and $v(n)$ represent $\varepsilon (n,n)$ and $v(n,n)$, respectively. Table \ref{Sum} summarizes the proposed algorithms, where $\delta$ is a small positive value, $\bm{I}$ is the identity matrix and $\tau $ is a small positive constant to avoid division by zero.
$\hfill\blacksquare$

\linespread{1.5}
\begin{table}[htbp]
    \caption{Summary of the proposed algorithms}
    \centering
    \begin{tabular}{l}
        \hline\hline
        Initialization: \\
        $\bm{w}(0)=\bm{0}$, $\bm{P}(0)=\delta \bm{I}$\\
        \hline
        Parameters: \\
        $\tau $, $\lambda $, $p$\\
        \hline
        Update: \\
        \textbf{for $n = 1, 2, 3, ...$}\\
        $v(n)=\frac{{{\left| e(n) \right|}^{p}}}{{{\left| e(n) \right|}^{2}}+\tau }$ (FxRLP)\\
        $v(n)=\frac{{{\log }^{p-1}}(1+\left| e(n) \right|)}{(1+\left| e(n) \right|)\left| e(n) \right|+\tau }$ (FxlogRLP)\\
        $\bm{K}(n)=\frac{v(n)\bm{P}(n-1){{\bm{x}}_{s}}(n)}{\lambda +v(n)\bm{x}_{s}^{T}(n)\bm{P}(n-1){{\bm{x}}_{s}}(n)}$\\
        $\bm{w}(n)=\bm{w}(n-1)+\bm{K}(n)e(n)$\\
        $\bm{P}(n)={{\lambda }^{-1}}\bm{P}(n-1)-{{\lambda }^{-1}}\bm{K}(n)\bm{x}_{s}^{T}(n)\bm{P}(n-1)$\\
        \textbf{end}\\
        \hline\hline
    \end{tabular}
    \label{Sum}
\end{table}

\noindent\emph{Remark 3:} As can be seen from Table \ref{Sum}, the FxRLP and FxlogRLP algorithms have the same structure except for the calculation of $v(n)$. Note that the FxRLP algorithm reduces to the FxRLS algorithm when $v(n)=1$ (i.e., $p=2$, $\tau =0$).
$\hfill\blacksquare$

\section{Mean stability analysis}

In this section, we perform the mean stability analyses of the proposed algorithms.

According to Table \ref{Sum}, the weight vector can be summarized and rewritten as follows
\begin{equation}
\bm{w}(n+1)=\bm{w}(n)+\frac{v(n)\bm{P}(n-1)\bm{x}_{s}(n)}{\lambda +v(n)\bm{x}_{s}^{T}(n)\bm{P}(n-1)\bm{x}_{s}(n)}e(n).
\label{weight}
\end{equation}

Define the weight deviation vector below
\begin{equation}
\bm{\tilde{w}}(n)={{\bm{w}}_{o}}-\bm{w}(n)
\end{equation}
where ${{\bm{w}}_{o}}$ denotes the optimal weight vector of the controller.

Subtracting (\ref{weight}) from ${{\bm{w}}_{o}}$, we have

\begin{equation}
\bm{\tilde{w}}(n+1)=\bm{\tilde{w}}(n)-\frac{v(n)\bm{P}(n-1)\bm{x}_{s}(n)}{\lambda +v(n)\bm{x}_{s}^{T}(n)\bm{P}(n-1)\bm{x}_{s}(n)}e(n).
\label{we}
\end{equation}

Taking the expectation on both sides of (\ref{we}) yields

\begin{equation}
\begin{aligned}
&E[\bm{\tilde{w}}(n+1)]\\
&=E[\bm{\tilde{w}}(n)]-E\left\{ \frac{v(n)\bm{P}(n-1)\bm{x}_{s}(n)}{\lambda +v(n)\bm{x}_{s}^{T}(n)\bm{P}(n-1)\bm{x}_{s}(n)}e(n) \right\}
\end{aligned}
\end{equation}
which can be expressed as
\begin{equation}
\begin{aligned}
&E[\bm{\tilde{w}}(n+1)]\\
&=E[\bm{\tilde{w}}(n)]-E\left[ \frac{v(n)\bm{P}(n-1)\bm{x}_{s}(n)\bm{x}_{s}^{T}(n)}{\lambda +v(n)\bm{x}_{s}^{T}(n)\bm{P}(n-1)\bm{x}_{s}(n)}\bm{\tilde{w}}(n) \right] \\
&\approx E[\bm{\tilde{w}}(n)]-E\left[ \frac{v(n)\bm{P}(n-1)\bm{x}_{s}(n)\bm{x}_{s}^{T}(n)}{\lambda +v(n)\bm{x}_{s}^{T}(n)\bm{P}(n-1)\bm{x}_{s}(n)} \right]E[\bm{\tilde{w}}(n)] \\
\end{aligned}
\end{equation}
where $E[\cdot]$ represents the expectation operator, and the approximation $ e(n)\approx \bm{x}_{s}^{T}(n)\bm{\tilde{w}}(n)$ was used.

Therefore, the algorithm converges in the mean if and only if
\begin{equation}
0<{{\lambda }_{\max }}\left\{ E\left[ \frac{v(n)\bm{P}(n-1)\bm{x}_{s}(n)\bm{x}_{s}^{T}(n)}{\lambda +v(n)\bm{x}_{s}^{T}(n)\bm{P}(n-1)\bm{x}_{s}(n)} \right] \right\}<2
\label{lammda}
\end{equation}
where ${{\lambda }_{\max }}\{\cdot \}$ denotes the largest eigenvalue of the matrix.

\noindent\emph{Remark 4:} Based on the fact that ${{\lambda }_{\max }}(\bm{AB})<\text{Tr}(\bm{AB})$ in (\ref{lammda}), it obtains
\begin{equation}
\begin{aligned}
&{{\lambda }_{\max }}\left\{ E\left[ \frac{v(n)\bm{P}(n-1)\bm{x}_{s}(n)\bm{x}_{s}^{T}(n)}{\lambda +v(n)\bm{x}_{s}^{T}(n)\bm{P}(n-1)\bm{x}_{s}(n)} \right] \right\}\\
&<E\left[ \frac{\text{Tr}(\bm{x}_{s}^{T}(n)\bm{P}(n-1)\bm{x}_{s}(n))}{\frac{\lambda}{v(n)} +\bm{x}_{s}^{T}(n)\bm{P}(n-1)\bm{x}_{s}(n)} \right]<1.
\end{aligned}
\end{equation}
Hence, the proposed algorithms are convergent in the mean if the input signal is persistently exciting.

\section{Simulation results}

\begin{figure} [htbp]
    \centering
    \includegraphics[scale=0.6] {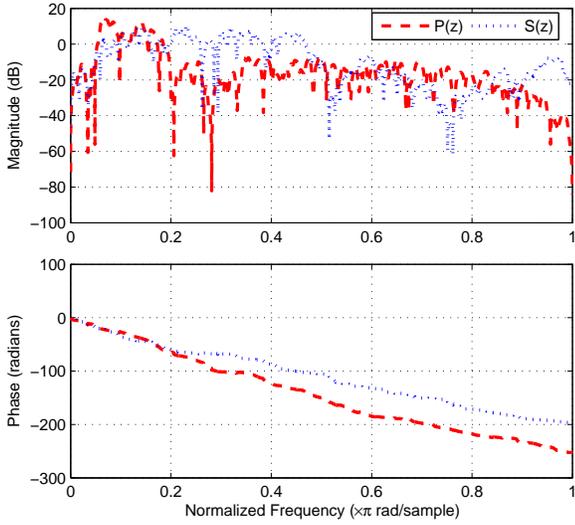}
    \caption{Magnitude and phase responses of the primary and secondary paths}
    \label{response}
\end{figure}

Simulations are carried out to examine the performance of the proposed algorithms for active impulsive noise control. The primary path $P(z)$ and secondary path $S(z)$ are modeled as finite impulse response (FIR) filters with the length of 256 and 100, respectively, and the estimated secondary path $\hat{S}(z)$ was exactly identified as $S(z)$. The magnitude and phase responses of the primary and secondary paths are shown in Fig. \ref{response}. The adaptive filter $W(z)$ is selected as an FIR filter with the length of 128. The reference signal $x(n)$ is generated through a standard symmetric $\alpha$-stable (S$\alpha$S) process with  $\alpha =1.35$ or $\alpha =1.55$ \cite{wu2015recursive}.

To evaluate the performance of the algorithms, the averaged noise reduction (ANR) is used:
\begin{equation}
\text{ANR}(n)=20\log \left( {{A}_{e}}(n)/{{A}_{d}}(n) \right),
\end{equation}
where
\begin{equation}
{{A}_{e}}(n)=\xi {{A}_{e}}(n-1)+(1-\xi )|e(n)|,
\end{equation}
\begin{equation}
{{A}_{d}}(n)=\xi {{A}_{d}}(n-1)+(1-\xi )|d(n)|,
\end{equation}
and $\xi =0.999$. The simulation results are obtained by ensemble averaging over 50 trials.

In the following simulations, the proposed algorithms are compared with the FxLMP \cite{leahy1995adaptive}, FxRLS \cite{kuo1996active}, and FxlogRLS \cite{wu2015recursive} algorithms, and the parameters are set as follows: the step size is $\mu=0.0001$ for the FxLMP algorithm; the forgetting factor $\lambda=0.999$ and $\delta=0.001$ are set for the FxRLS, FxlogRLS, FxRLP and FxlogRLP algorithms; $p=1.3(1.5)$ is set for the FxLMP, FxRLP and FxlogRLP algorithms under the S$\alpha$S signal with  $\alpha =1.35(1.55)$; $\tau=0.001$ is set for the FxRLP and FxlogRLP algorithms.

\begin{figure} [htbp]
    \centering
    \includegraphics[scale=0.6] {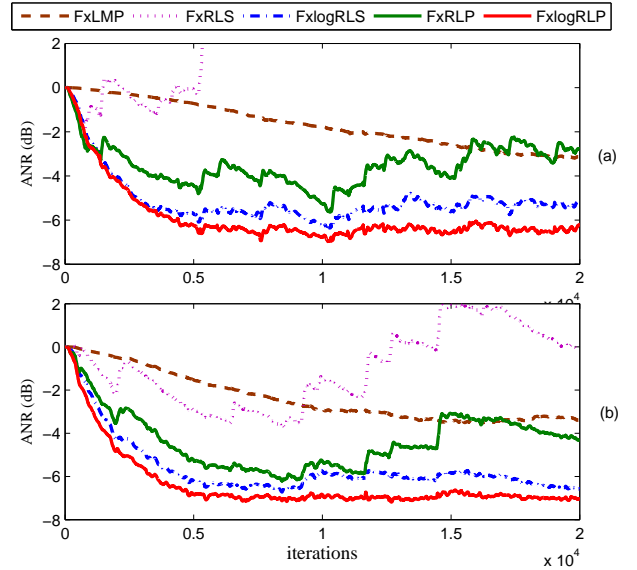}
    \caption{ANR curves of algorithms. (a) $\alpha =1.35$; (b) $\alpha =1.55$.}
    \label{Compare}
\end{figure}

The performance comparison of the algorithms is shown in Fig. \ref{Compare}. As can be seen, the FxRLS algorithm diverges because the second-order moment of the target noise is not finite. The FxRLP algorithm shows faster convergence speed relative to the FxLMP algorithm, but has a higher steady-state ANR as compared to the FxlogRLS algorithm. Owing to the new cost function which integrates the $p$-order moment with the logarithmic-order moment, the FxlogRLP algorithm presents the best performance among them.

\section{Conclusion}
By minimizing the weighted summation of the $p$-power of the \emph{a posteriori} errors, we proposed a FxRLP algorithm for active impulsive noise control. Moreover, by integrating the $p$-order moment with the logarithmic-order moment, a FxlogRLP algorithm was developed. Simulation results confirmed that the FxlogRLP algorithm outperforms the state-of-the-art algorithms.

\bibliographystyle{IEEEtran}
\bibliography{IEEEabrv,mybibfile}

\end{document}